\documentclass[fleqn,12pt,twoside]{article}
\usepackage{espcrc1}
\usepackage{epsfig}
\usepackage{graphicx}
\title{Source Parameters from Identified Hadron Spectra and HBT Radii for Au-Au Collisions at $\sqrt{s_{NN}} = $200 GeV in PHENIX}
\author{J. M. Burward-Hoy\address[LLNL]{Physics and Applied Technology Directorate, Lawrence Livermore National Laboratory, 7000 East Avenue L-305, Livermore, California  94550}, for the PHENIX Collaboration\footnote{For the full PHENIX Collaboration author list and acknowledgements, see Appendix ``Collaborations'' of this volume.}}
\begin{document}
\maketitle
\begin{abstract}
The characteristics of the particle emitting source  are deduced from
low $p_T$ identified hadron spectra  ($(m_T-m_0) < 1$ GeV) and HBT
radii using a hydrodynamic interpretation.  From the most peripheral
to the most central data, the single particle spectra are fit
simultaneously for all $\pi^{\pm}$, $K^{\pm}$, and $\overline{p}/p$
using the parameterization in \cite{ref1} and assuming a linear
transverse flow profile.  Within the systematic uncertainties, the
expansion parameters $T_{fo}$ and $\beta_{T}$,  respectively decrease
and increase with the number of  participants, saturating for both at
mid-centrality.  The expansion using analytic calculations of the
$k_T$ dependence of HBT radii in \cite{ref2} is fit to the data but no
$\chi^{2}$ minimum is found.
\end{abstract}
\section{INTRODUCTION}
Identified charged hadrons in 11 different centrality selections
\cite{ref3} and the transverse momentum ($p_T$) dependence of HBT
radii in 9 $k_T$ bins \cite{ref4}  are measured in Au-Au collisions
at 200 GeV by the PHENIX Experiment \cite{ref5}.   In both the 200 GeV
and previously measured 130 GeV data \cite{ref6}, the  $\langle p_{T} \rangle$ of
all particles increases from the most peripheral to the most central
events and with heavier particle mass ($m_0$).  The dependence of the
$\langle p_{T} \rangle$ on $m_0$ suggests a radial expansion,  and its
dependence on the number of participant nucleons ($N_{part}$) may be
due to an increasing radial expansion from  peripheral to central
events.  The $k_T$ dependence of the HBT radii  was also observed and
interpreted as a radial expansion.

Both the spectra and the $k_T$ dependence of the HBT radii are 
fit using parameterizations based on a simple model for the 
source, where fluid elements are each in
local thermal equilibrium and move in space-time with a hydrodynamic
expansion \cite{ref1,ref2}.  The assumptions are:  $(1)$ no
temperature  gradients, $(2)$ longitudinal boost invariance along
the collision axis z,  $(3)$ infinite extent in space-time rapidity $\eta$, and $(4)$
cylindrical symmetry  with radius r.  The particles are emitted along
a hyperbola of constant  proper time $\tau_{0} = \sqrt{t^2 - z^2}$ and
short emission  duration, $\Delta t < $ 1 fm/c.

The $m_T$ dependence of the yield $\frac{d^2N}{m_t dm_t dy}|_{y=0}$ is  calculated
after integrating the source over space-time (azimuthal and rapidity
coordinates) \cite{ref1}.  It is assumed that all particles decouple
kinematically on the freeze-out hypersurface  at the  same freeze-out
temperature $T_{fo}$, and that the particles collectively expand with
a velocity profile $\beta_T \left( r \right) = \beta_T r/R$ where $R$
is the geometric radius, $r$ is the transverse coordinate, and
$\beta_T$ is the  surface velocity.  (For a box profile, the average
velocity is  $\langle \beta_T \rangle$ $= 2\beta_T/3$ \cite{ref7}).
The particle density distribution is assumed to be independent of the
radial position in the fits to the single particle spectra.  In the
previous 130 GeV analysis \cite{ref8}, a Gaussian density profile
increases $\beta_T$ by $\approx$ 2\% with a negligible difference in
$T_{fo}$, while a parabolic velocity profile increases $\beta_T$ by
13\% and $T_{fo}$ by 5\%.

We use analytic expressions to calculate the HBT radii \cite{ref2}.  A
linear flow rapidity profile in the transverse plane is assumed and a
Gaussian distribution is used for the particle density dependence on
r.  The parameters are the geometric radius R, the freeze-out
temperature T, the flow rapidity at the surface $\eta_T$ ($\beta_T$
$= \rm{tanh} \left( \eta_T \right)$) and the freeze-out proper time
$\tau_0$.
\section{RESULTS}
\subsection{Fitting the single particle spectra}
\begin{figure}[tb]
\vspace{-0.3in}
\begin{minipage}[htb]{70mm}
\includegraphics[scale=0.40]{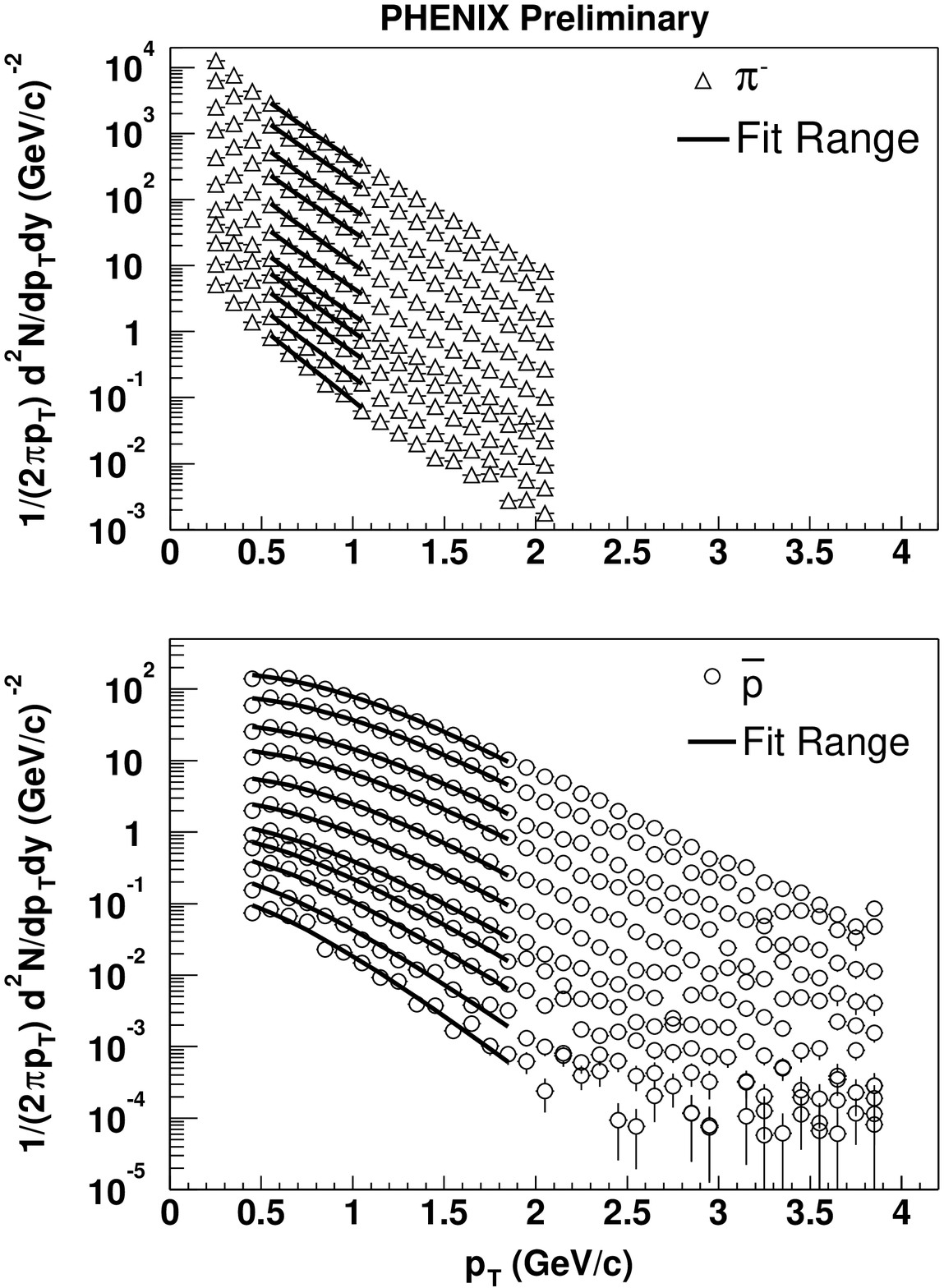}
\vspace{-0.6in}
\caption{Simultaneous fits in the range $(m_T-m_0)<1$ (solid lines)
for $\pi^{-}$ (top) and $\overline{p}$ (bottom) in all 11 centralities
(scaled for visual  clarity) \cite{ref4}.  The $\pi$ resonance region
is excluded in the fit.  }
\label{fig1}
\end{minipage}
\hspace{\fill}
\vspace{-0.3in}
\hspace{\fill}
\begin{minipage}[tb]{80mm}
\vspace{-0.3in}
\includegraphics[scale=0.40]{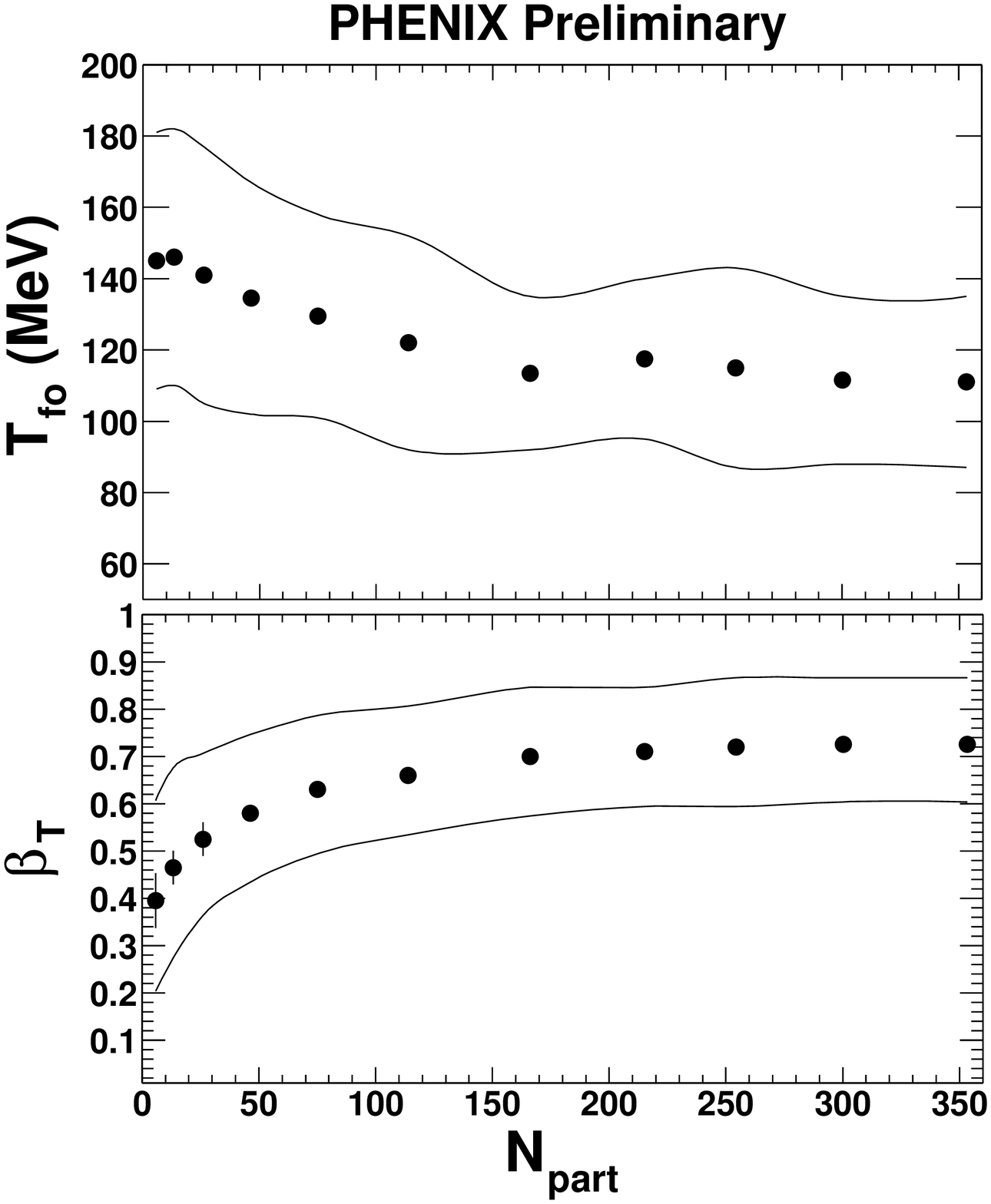}
\vspace{-0.7in}
\caption{The expansion in each centrality.  The top panel is $T_{fo}$
and the bottom is $\beta_T$, both plotted as a function of
$N_{part}$. }
\label{fig2}
\end{minipage}
\vspace{-0.05in}
\end{figure}

In order to minimize contributions from hard processes, all $m_T$
dependent particle yields are fit in the range $(m_T-m_0) < 1$ GeV.
As resonance decays are known to produce pions at low $p_T$
\cite{ref9}, we place a lower $p_T$ threshold of 500 MeV/c on $\pi$ in
the fit.  A similar  approach was followed by NA44, E814, and other
experiments at lower energies.   In Fig.~\ref{fig1}, $\pi^{-}$ and
$\overline{p}$ yields are shown as a function of $p_T$ for each event
centrality \cite{ref4}.  The top 5 centralities are scaled for visual
clarity.  The solid lines are the simultaneous fits in the limited
$p_T$ range.  Similar results are obtained for $K^{\pm}$, $\pi^{+}$,
and p yields.

The systematic uncertainty in $T_{fo}$ is determined by adding in
quadrature the change in inverse slope due to the $p_T$ dependent
uncertainties in each particle yield at low $p_T$.  For $\pi^{\pm}$,
$K^{\pm}$, and $\overline{p}/p$, the uncertainty is $\pm10$, $\pm13$,
and $16$ MeV respectively.  Added in quadrature, the total systematic
uncertainty in the inverse  slope is $\pm23$ MeV.  The systematic
uncertainty in $\beta_T$ is dominated by the uncertainty in the
$\overline{p}/p$  spectral shape at low $p_T$ and is determined by
measuring the change in  $\beta_T$ after fitting for $p_T>0.85$ GeV/c.  
The systematic uncertainty in $\beta_T$ is $17.5\%$.

For the 5\% most central events,  particles are coupled to an expanding
system with a surface velocity of $\beta_T = 0.7\pm0.2 (syst.)$ and
decouple at a common temperature of $T_{fo} = 110\pm23 (syst.)$ MeV 
with negligible statistical errors.  For the most peripheral
events, $T_{fo} = 135\pm3 (stat.) \pm23 (syst.)$  and $\beta_T =
0.46\pm0.02 (stat.)\pm0.2 (syst.)$.  The statistical
error only is included in the fit, resulting in  $\chi^{2}/dof =
260.9/52$ for the most central and $321.5/52$ for the most peripheral
events.  At 130 GeV, similar results were obtained, with  $\beta_T =
0.70\pm0.01$, $T_{fo} = 121\pm4$, and $\chi^{2}/dof = 34.0/40.0$  for
the most central events (statistical and systematic errors are added
in quadrature before the fit) \cite{ref10}.

The fit results of all particles within each event centrality are
shown in Fig.~\ref{fig2}.   The top panel is $T_{fo}$ and the bottom
panel is $\beta_T$, both plotted as a function of $N_{part}$.  Within
the systematic uncertainties, the expansion parameters respectively
decrease and increase with the number of participants, saturating at
mid-centrality.

\subsection{Fitting the $k_T$ dependence of the HBT radii}
\begin{figure}[tb]
\begin{minipage}[b]{0.6\linewidth}
\begin{center}
\includegraphics[scale=0.5]{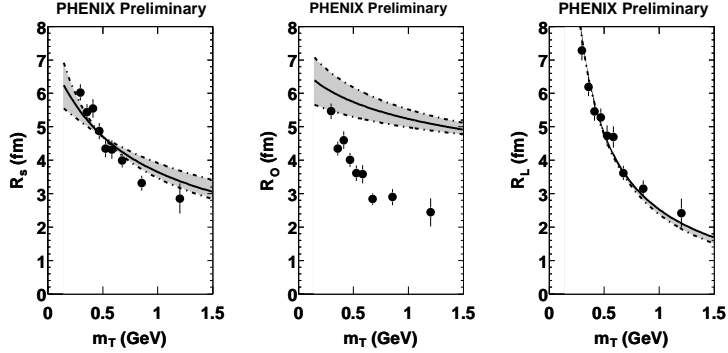}
\end{center}
\vspace{-0.4in}
\end{minipage}
\hfill
\parbox[b]{0.35\textwidth}{\sloppy
\caption{Constraining fits to 2$\pi^{-}$ HBT radii in 10\% central
events using the expansion measured from the spectra.  The shaded
region is the systematic uncertainty from $\beta_T$ and $T_{fo}$.}
\protect\label{fig3}}
\end{figure}

The HBT radii are measured from identical charged $\pi$ pairs in 9
$k_T$ selections for 10\% central events \cite{ref5}.  The systematic 
uncertainty in the data is 8.2\%, 16.1\%, and 8.3\% for $R_{\rm{s}}$,
$R_{\rm{o}}$, and $R_{\rm{L}}$, respectively.  A simultaneous fit to
the HBT data could not be found over a broad range of parameter space.
As an example, if the parameters $\beta_T$ and $T_{fo}$ are set to the 
values from the spectra analysis, then the fit to the HBT results 
constrains R and $\tau_{0}$ from the $R_{\rm{s}}$ and $R_{\rm{L}}$ data 
respectively, yet the model overpredicts $R_{o}$ by more than 
3$\sigma$ for all but the first $m_T$ data point (Fig.~\ref{fig3}).
The systematic uncertainties in $\beta_T$ and  $T_{fo}$ are 
represented by the shaded region.  Within these boundaries, R ranges 
between $6.9-16.8$ fm and $\tau_{0}$ ranges  between $11.2-16.7$ fm/c.

\begin{figure}[tb]
\begin{minipage}[b]{0.6\linewidth}
\includegraphics[scale=0.5]{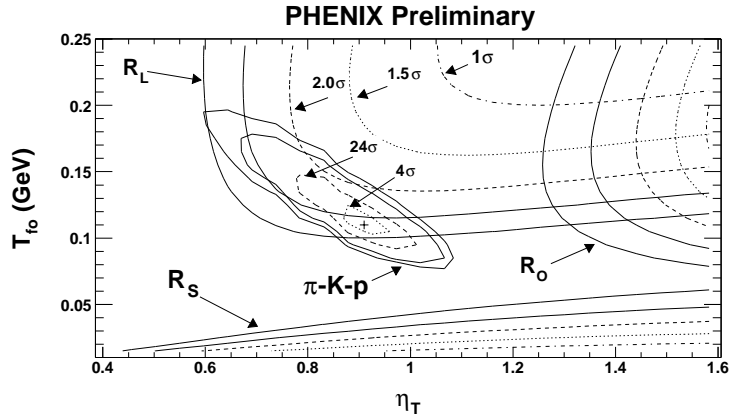}
\vspace{-0.6in}
\end{minipage}
\hfill
\parbox[b]{0.35\textwidth}{\sloppy
\caption{The $\chi^{2}$ contour levels for the expansion parameters
$T_{\rm{fo}}$ (vertical) and $\eta_{\rm{T}}$ (horizontal) after
fitting $\pi^{\pm}$, $K^{\pm}$, $\overline{p}/p$ spectra and 2$\pi$
HBT radii as indicated.  Both scales are zero suppressed.}
\protect\label{fig4}}
\end{figure}

The $\chi^{2}$ contour levels of the expansion parameters
$T_{\rm{fo}}$ (vertical) and $\eta_{\rm{T}}$ (horizontal) are shown
for simultaneous fits to the spectra and separate fits to each HBT 
radius in Fig.~\ref{fig4}.  We note that no $\chi^{2}$ minima are 
found, hence the contours are  not closed.  For the spectra, the 
contours are closed and show an anticorrelation, however there is 
no overlap with the HBT contours.  The HBT radius $R_{\rm{s}}$ 
prefers large flow rapidity $\eta_{\rm{T}}>1.0$ and low 
temperatures $T_{\rm{fo}}<50$ MeV.  The parameterization has the 
most difficulty reproducing $R_{\rm{o}}$.
\section{CONCLUSION}
The single particle spectra are qualitatively described by a
hydrodynamic parameterization that assumes boost invariance and a
linear transverse flow profile.  The transverse expansion in 11
different centrality classes is extracted from the single particle
spectra.  Within the systematic uncertainties, the expansion
parameters $T_{\rm{fo}}$ and $\beta_{\rm{T}}$, respectively decrease
and increase with the number of participants, saturating at
mid-centrality.  Expressions for the HBT radii based on similar
hydrodynamic assumptions and Gaussian density profiles do not describe
the identical $\pi$ pair data.  Such fits worked well at CERN SPS
energies \cite{ref11}, but fail at RHIC energies.


\begin{thebibliography}{9}
\bibitem{ref1} Schnedermann, Sollfrank, and Heinz, Phys. Rev. C {\bf 48} (1993) 2462.
\bibitem{ref2} Wiedemann, Scotto, and Heinz, Phys. Rev. C {\bf 53} (1996) 918.
\bibitem{ref3} T. Chujo, in these Proceedings.
\bibitem{ref4} A. Enokizono, in these Proceedings.
\bibitem{ref5}D. Morrison et al., Nucl. Phys. A {\bf 638} (1998); W. Zajc, Nucl. Phys. A {\bf 698} (2002) 39.; K. Adcox et al., accepted for publication in Nucl. Instr. Meth. A Special issue.
\bibitem{ref6} K. Adcox et al., Phys. Rev. Lett. {\bf 88} (2002) 242301.
\bibitem{ref7} S. Esumi, S. Chapman, H. van Hecke, and N. Xu, Phys. Rev. C {\bf 55} (1997).
\bibitem{ref8} J. Burward-Hoy for the PHENIX Collaboration, Proceedings of the 18th Winter Workshop on Nuclear Dynamics, nucl-ex/0206016. 
\bibitem{ref9} J. Sollfrank, P. Koch, and U. Heinz, Phys. Lett. B {\bf 252} (1990) 256; H. Boggild et al., Z. Phys. C {\bf 69} (1996) 621; J. Barette et al., Phys. Lett. B {\bf 351} (1995) 93.
\bibitem{ref10} K. Adcox et al., to be submitted to Phys. Rev. C (2002).
\bibitem{ref11} A. Appelshauser et al., Eur. Phys. J. C {\bf 2} (1998) 661.
\end{thebibliography}
\end{document}